\documentstyle[twoside,fleqn,espcrc2,epsf]{article}

\newcommand{\AmS}{{\protect\the\textfont2
  A\kern-.1667em\lower.5ex\hbox{M}\kern-.125emS}}

\newcommand{\be}{\begin{equation}}
\newcommand{\ee}{\end{equation}}
\newcommand{\bea}{\begin{eqnarray}}
\newcommand{\eea}{\end{eqnarray}}

\newcommand{\RC}{{\bf RC}}
\newcommand{\qRC}{{\bf qRC}}
\newcommand{\qPW}{{\bf qPW}}

\title{
\vspace*{-44pt}
{\normalsize \hfill {\sf UTCCP-P-70}} \\
\vspace*{-5pt}
{\normalsize \hfill {\sf UTHEP-410}} \\
\vspace*{-5pt}
{\normalsize \hfill {\sf Sept.\ 1999}} \\
Light hadron spectrum and quark masses in QCD 
with two flavors of dynamical quarks
\thanks{Talk presented by T.~Kaneko}
}

\author{CP-PACS Collaboration :\\
        A.~Ali Khan\address{Center for Computational Physics, University of Tsukuba, Tsukuba, Ibaraki 305-8577, Japan},
        S.~Aoki\address{Institute of Physics, University of Tsukuba, Tsukuba, Ibaraki 305-8571, Japan},
        G.~Boyd$^{\rm a}$,
        R.~Burkhalter$^{\rm a,b}$,
        S.~Ejiri$^{\rm a}$,
        M.~Fukugita\address{Institute for Cosmic Ray Research,
        University of Tokyo, Tanashi, Tokyo 188-8502, Japan},
        S.~Hashimoto\address{High Energy Accelerator Research Organization
        (KEK), Tsukuba, Ibaraki 305-0801, Japan},
        N.~Ishizuka$^{\rm a,b}$,
        Y.~Iwasaki$^{\rm a,b}$,
        K.~Kanaya$^{\rm a,b}$,
        T.~Kaneko$^{\rm a}$,
        Y.~Kuramashi$^{\rm d}$,
        T.~Manke$^{\rm a}$,
        K.~Nagai$^{\rm a}$,
        M.~Okawa$^{\rm d}$,
        H.P.~Shanahan\address{DAMTP, University of Cambridge, 
        Cambridge, CB3 9EW, England, UK},
        A.~Ukawa$^{\rm a,b}$, and
        T.~Yoshi\'e$^{\rm a,b}$ }

\begin{document}

\begin{abstract}
We present updated results of the CP-PACS calculation
of the light hadron spectrum in $N_{\rm f}\!=\!2$ full QCD.
Simulations are made with an RG-improved gauge
action and a tadpole-improved clover quark action for sea quark
masses corresponding to $m_{\rm PS}/m_{\rm V} \approx 0.8$--0.6 and 
the lattice spacing $a=0.22$--0.09 fm.
A comparison of the  $N_{\rm f}\!=\!2$ QCD spectrum with  new quenched results, obtained
with the same improved action, shows clearly the existence of sea quark effects 
in vector meson masses.  Results for light quark masses are also presented.
\end{abstract}

\maketitle
\setcounter{footnote}{0}
\section{Introduction}

Understanding sea quark effects in the light hadron spectrum is 
an important issue, sharpened by the recent finding of a systematic deviation 
of the quenched spectrum from experiment\cite{CP-PACS.Quenched}.  
To this end, we have been pursuing  $N_{\rm f}\!=\!2$  QCD simulations 
using an RG-improved gauge action and a tadpole-improved clover quark 
action~\cite{Ruedi.Review}, to be called \RC\ simulations in this article. 

The parameters of these simulations are listed in Table~\ref{tab:param}.
The statistics at $\beta \! = \! 2.2$ have been increased since Lattice'98, 
and the runs at $\beta=2.1$ are new.
In addition we have carried out quenched simulations with the same
improved action, referred to as \qRC, for a direct 
comparison of the full and quenched spectrum.
The $\beta$ values of these runs, given in Table~\ref{tab:param},
are chosen so that the lattice spacing fixed by
the string tension matches that of full QCD 
for each value of sea quark mass 
at $\beta\! =\! 1.95$ and 2.1.  
Quenched hadron masses are calculated for valence quark masses such that 
$m_{\rm PS}/m_{\rm V} \approx $ 0.8--0.5, which is similar to those 
in the \RC\ runs. 

In this report we present updated results of the full QCD spectrum and 
light quark masses.  We also discuss sea quark effects by
comparing the \RC\ and \qRC\ results. 
For reference we use quenched results with 
the plaquette gauge and Wilson quark action~\cite{CP-PACS.Quenched} as well, 
which we denote as \qPW .

\begin{table}[t]
\vspace{-3mm}
\setlength{\tabcolsep}{0.2pc}
\caption{
          Parameters in \RC\ and \qRC\ simulations. 
         Scale $a_{\sigma}$ is fixed by $\protect \sqrt{\sigma}=440$~MeV.  
         \qRC\ runs have 200 configurations for each $\beta$.
}
\vspace{+5pt}
\label{tab:param}
\begin{tabular}{lllll}
 \multicolumn{5}{l}{\RC\ simulations} \\
\hline
 lattice   & $K_{\rm sea}$ & \#traj. & $m_\pi/m_\rho$ & $a_\sigma$~[fm]\\
\hline
$12^3{\times}24$     &  0.1409   & 6250 &   0.806(1) & 0.289(3) \\  
$\beta\!=\!1.80$     &  0.1430   & 5000 &   0.753(1) & 0.152(2) \\
$c_{SW}\!=\!1.60$    &  0.1445   & 7000 &   0.696(2) & 0.269(3) \\
$a\!=\!0.215(2)$~fm  &  0.1464   & 5250 &   0.548(4) & 0.248(2) \\
\hline
$16^3{\times}32$     &  0.1375   & 7000 &   0.805(1) & 0.204(1) \\
$\beta\!=\!1.95$     &  0.1390   & 7000 &   0.751(1) & 0.193(2) \\
$c_{SW}\!=\!1.53$    &  0.1400   & 7000 &   0.688(1) & 0.181(1) \\
$a\!=\!0.153(2)$~fm  &  0.1410   & 7000 &   0.586(3) & 0.170(1) \\
                  \cline{2-4}
\hline
$24^3{\times}48$     &  0.1357   & 2000 &   0.806(2) & 0.1342(6) \\ 
$\beta\!=\!2.10$     &  0.1367   & 2000 &   0.757(2) & 0.1259(5) \\
$c_{SW}\!=\!1.47$    &  0.1374   & 2000 &   0.690(3) & 0.1201(5) \\
$a\!=\!0.108(2)$~fm  &  0.1382   & 2000 &   0.575(6) & 0.1128(3) \\
                  \cline{2-4}
\hline
$24^3{\times}48$     &  0.1351   & 2000 &   0.800(2) & 0.1049(2) \\ 
$\beta\!=\!2.20$     &  0.1358   & 2000 &   0.754(2) & 0.1012(3) \\
$c_{SW}\!=\!1.44$    &  0.1363   & 2000 &   0.704(3) & 0.0977(3) \\
$a\!=\!0.086(3)$~fm  &  0.1368   & 2000 &   0.629(5) & 0.0947(2) \\
                  \cline{2-4}
\hline
 \end{tabular}
 \vspace{+5pt}

 \begin{tabular}{lllll}
 \multicolumn{5}{l}{\qRC\ simulations} \\
 \hline
 \multicolumn{2}{l}{$16^3{\times}32$} &  &  \multicolumn{2}{l}{$24^3{\times}48$}\\
 \cline{1-2} \cline{4-5}
 $\beta$   &   $a_\sigma$~[fm]   & &  $\beta$   &  $a_\sigma$~[fm]   \\
 \cline{1-2} \cline{4-5}
 2.187     &  0.2079(15)  & &  2.416     &  0.1359(7)  \\  
 2.214     &  0.1977(13)  & &  2.456     &  0.1266(13) \\
 2.247     &  0.1853(9)   & &  2.487     &  0.1206(9)  \\
 2.281     &  0.1727(10)  & &  2.528     &  0.1130(9)  \\
 2.334     &  0.1577(9)   & &  2.575     &  0.1065(7)  \\
 \hline
\end{tabular}
\vspace{-15pt}
\end{table}

\section{Full QCD spectrum}
The analysis procedure of our full QCD spectrum data follows 
that in Ref.~\cite{Ruedi.Review}: $m_\pi$ and $m_\rho$ are used to set the
scale and determine the up and down quark mass $m_{ud}$, while 
the strange quark mass $m_s$ is fixed from either $m_K$ or $m_\phi$. 
We tested several fitting forms for the continuum extrapolation, and 
found that the fit is stable; e.g., for the meson masses, 
linear extrapolations in $a$ and in $a\alpha_{\overline{\rm MS}}$ 
are consistent with each other
and a quadratic fit in $a$ is also
consistent within 2 standard deviations.
Here, we present results from the linear extrapolation in $a$.

\begin{figure}[tb]
\vspace{-13mm}
\begin{center}
\leavevmode
\epsfxsize=7.5cm
\epsfbox{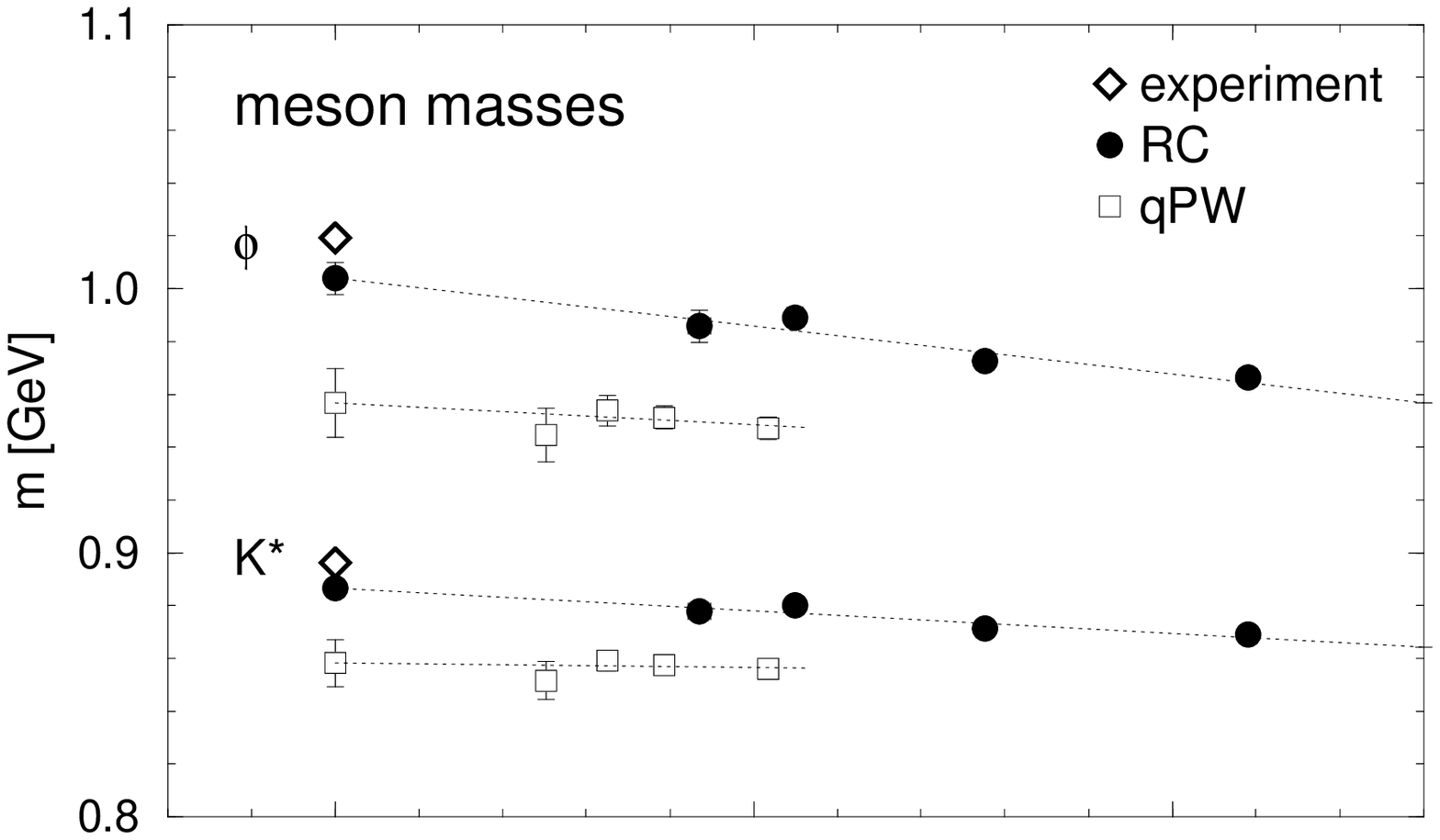}
\vspace{-26mm}
\leavevmode
\epsfxsize=7.5cm
\epsfbox{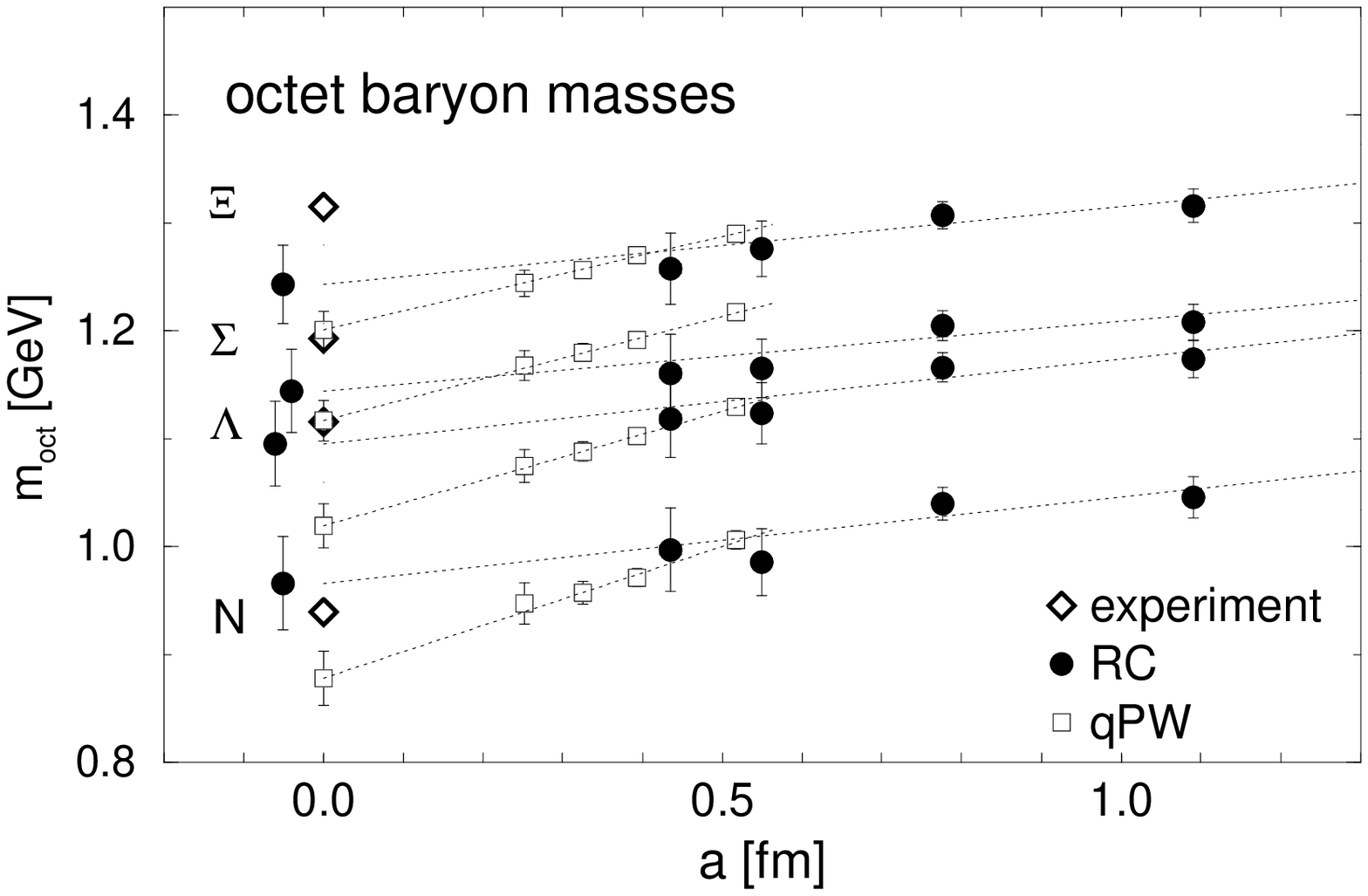}
\end{center}
\vspace{-30mm}
\caption{
Typical hadron masses with $m_K$-input.
}
\label{fig:spectrum}
\vspace{-19pt}
\end{figure}

Fig.~\ref{fig:spectrum} shows an update of results for vector
meson and octet baryon masses in comparison to those from the 
\qPW\ simulation. With increased statistics at $\beta\! =\!
2.2$ and new points at $\beta\!=\!2.1$, 
we find our conclusion to remain unchanged since Lattice'98, 
{\it i.e.,} meson masses in full QCD extrapolate significantly closer to
experiment than in quenched QCD. For baryons, the statistical errors are
still too large to draw definitive conclusions.

\section{Sea quark mass dependence}
In order to obtain a deeper understanding of 
the sea quark effect in meson masses, 
we investigate how their values depend on the sea 
quark mass.  In this test, the valence strange quark mass is fixed by 
a phenomenological value of the ratio
$m_{\eta_{\overline{s}s}}/m_{\phi}=0.674$.
To avoid uncertainties that may arise from chiral extrapolations, 
the light dynamical quark mass is set to one of the values corresponding 
to $m_{\rm PS}/m_{\rm V} = 0.7, 0.6$ or 0.5. 
The values of the masses ``$m_{K^*}$'' and ``$m_\rho$'' of fictitious mesons  
for such quark masses can then be determined by interpolations or 
short extrapolations of hadron mass results. 

\begin{figure}[tb]
\vspace{-12mm}
\begin{center}
\leavevmode
\epsfxsize=7.5cm
\epsfbox{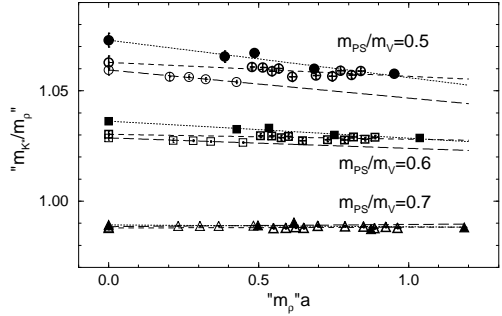}
\end{center}
\vspace{-29mm}
\caption{
Fictitious mass ratio ``$m_{K^*}/m_{\rho}$'' defined in the text 
at different sea quark masses.
Filled, thick open and thin open symbols are the results
from \RC\ , \qRC\ and \qPW\ simulations, respectively.
}
\label{fig:massratio}
\vspace{-19pt}
\end{figure}

In Fig.~\ref{fig:massratio}, we plot ``$m_{K^*}/m_{\rho}$'' as a 
function of the lattice spacing normalized by ``$m_\rho$'' for different 
sea quark masses.
Making linear extrapolations in $a$, we observe that the continuum limits of 
the two quenched simulations \qRC\ and \qPW\ are consistent.  
On the other hand, the full QCD result from \RC\ exhibits an increasingly 
clearer deviation from the quenched value toward lighter sea quark masses.  
We consider that this result provides a clear demonstration of 
the sea quark effect on vector meson masses.

\section{Quark masses}

\begin{figure}[tb]
\vspace{-9mm}
\begin{center}
\leavevmode
\epsfxsize=7.5cm
\epsfbox{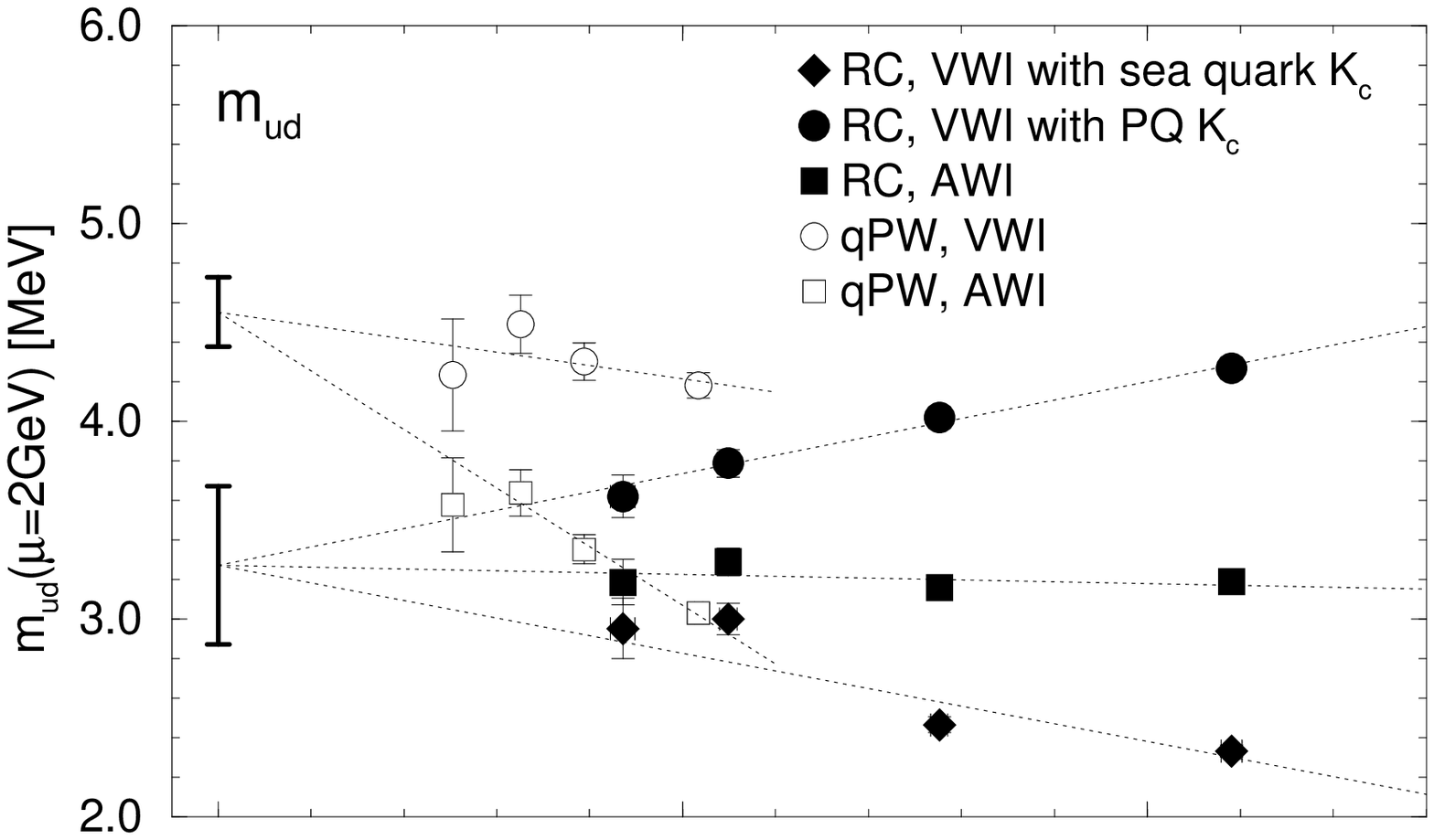}
\vspace{-26mm}
\leavevmode
\epsfxsize=7.5cm
\epsfbox{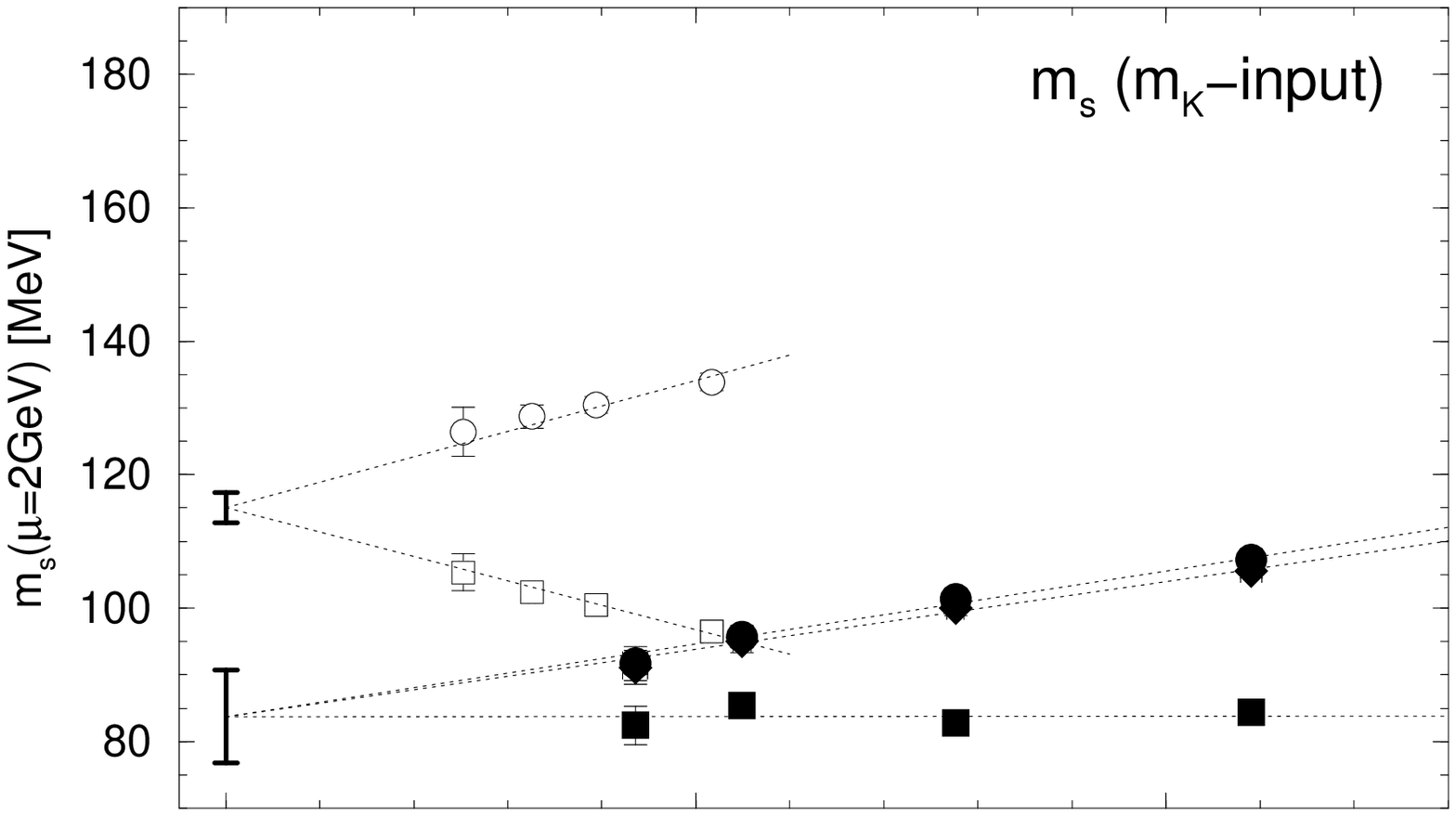}
\vspace{-26mm}
\leavevmode
\epsfxsize=7.5cm
\epsfbox{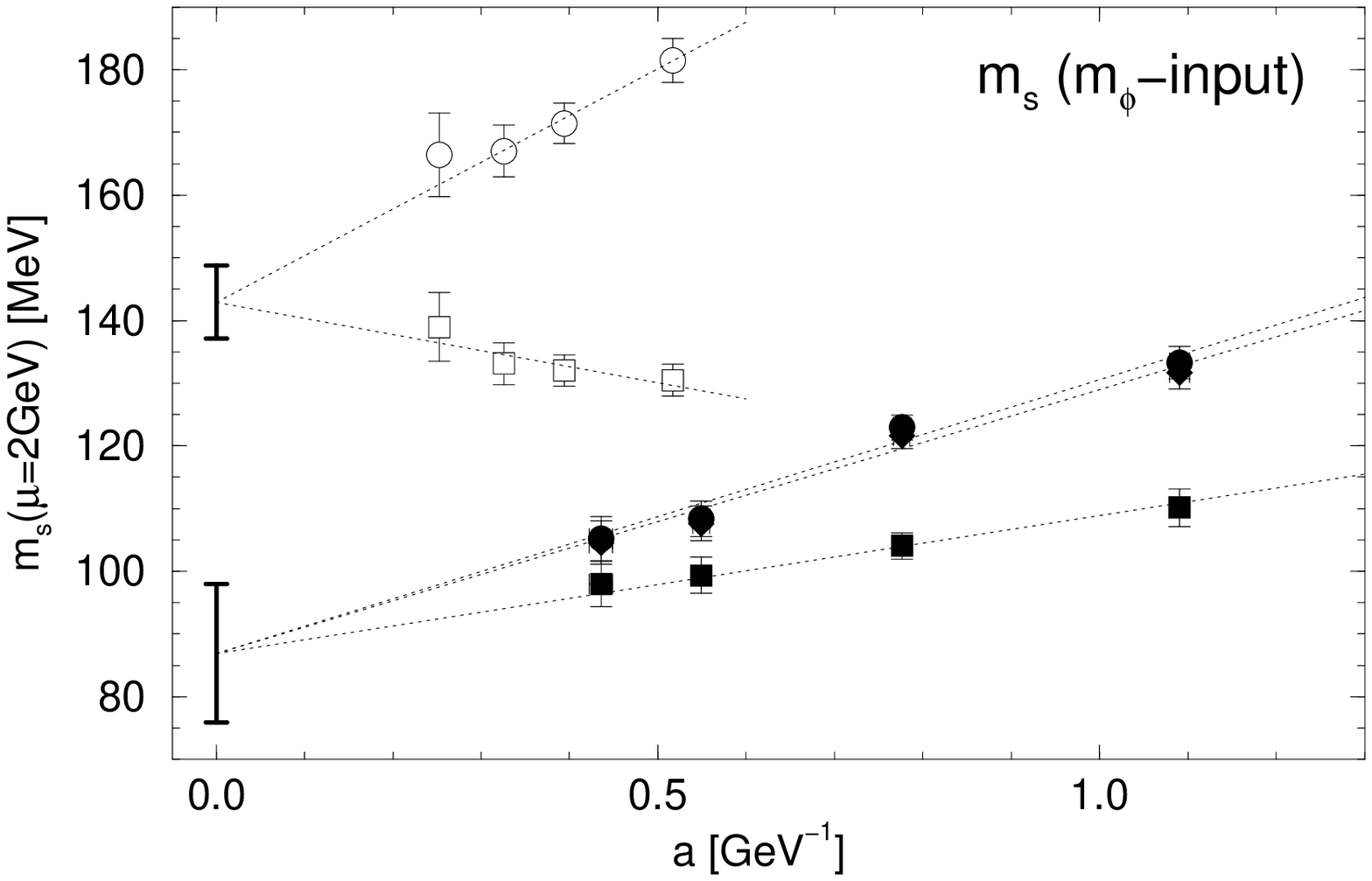}
\end{center}
\vspace{-30mm}
\caption{
Continuum extrapolation of $m_{ud}$ (top) 
and $m_s$ with $m_{K}$-input(middle) and $m_{\phi}$-input(bottom).
}
\label{fig:mq}
\vspace{-10pt}
\vspace{-2mm}
\end{figure}

We plot our results for light quark masses 
in the $\overline{\rm MS}$ scheme at $\mu=$2~GeV in Fig.~\ref{fig:mq}, 
together with the quenched results of Ref.~\cite{CP-PACS.Quenched}.
Continuum extrapolations are made linearly in $a$ with the constraint 
that the three definitions (using axial vector Ward identity(AWI) or 
vector Ward identity(VWI) with either $K_c$ from 
sea quarks or partially quenched $K_c$ ) yield the same value.  
We confirm our previous finding\cite{Ruedi.Review} 
that i) quark masses in full QCD are 
much smaller than those in quenched QCD, and ii) the large
discrepancy in the strange quark mass determined from $m_K$ or
$m_\phi$, observed in quenched QCD, is much reduced.

Our current estimate for quark masses in $N_{\rm f} \!=\! 2$ QCD are
$m_{ud}=3.3(4)$~MeV, $m_s =84(7)$~MeV ($K$-input) and $m_s
=87(11)$~MeV ($\phi$-input). 
The quoted errors include our estimate of the systematic errors due to the
choice of functional form of continuum extrapolations 
and the definition of the $\overline{\rm MS}$ coupling used in
the one-loop tadpole improved renormalization factor.
%
%

Our results for quark masses are smaller than the values often used 
in phenomenology\cite{ChPT.SumRules}, 
though the ratio $m_{ud}/m_{s} =$ 26(3) is consistent 
with the result 24.4(1.5)\cite{ChPT.Quarkmass} from chiral perturbation 
theory. 
The small values are quite interesting, especially for the strange 
quark mass; a smaller strange quark mass raises the prediction
of the Standard Model for the direct CP violation parameter 
${\rm Re}(\epsilon'/\epsilon)$, 
as strongly favored by the experimental results from the KTeV~\cite{ktev} 
and NA31 Collaborations~\cite{dCPviolation}.  

\vspace{10pt}

This work is supported in part by Grants-in-Aid 
of~the~Ministry~of~Education~(Nos.~09304029,
10640246,~10640248,~11640250,~11640294, 10740107,  11740162). 
SE and KN are JSPS Research Fellows.  AAK, TM and HPS are
supported by the Research for the Future Program of JSPS,
and HPS also by the Leverhulme foundation.

\end{document}